\documentclass[article,12pt,authoryear]{elsarticle}
\journal{Astronomy and Computing}
\usepackage{amssymb}
\usepackage{amsmath}
\usepackage{lineno}
\usepackage{url}
\usepackage[letterpaper, margin=1in]{geometry}
\begin{document}

\begin{frontmatter}

\title{From Global Climate Models (GCMs) to Exoplanet Spectra with the Global Emission Spectra (GlobES)}

\author[label1,label2,label3]{Thomas J. Fauchez}
\author[label1,label3]{Geronimo L. Villanueva}
\author[label1,label2,label3]{Vincent Kofman}
\author[label4]{Gabriella Suissa}
\author[label1]{Ravi K. Kopparapu}

\affiliation[label1]{organization={NASA Goddard Space Flight Center},
            addressline={8800 Greenbelt Road},
            city={Greenbelt},
            postcode={20771},
            state={MD},
            country={USA}}
 \affiliation[label2]{organization={Integrated Space Science and Technology Institute, Department of Physics, American University},
            addressline={ 4400 Massachusetts Ave NW},
            city={Washington},
            postcode={20016},
            state={DC},
            country={USA}}           
\affiliation[label3]{organization={Sellers Exoplanet Environments Collaboration},
            addressline={8800 Greenbelt Road},
            city={Greenbelt},
            postcode={20771},
            state={MD},
            country={USA}}
\affiliation[label4]{organization={University of Washington},
            addressline={1410 NE Campus Pkwy},
            city={Seattle},
            postcode={98195},
            state={WA},
            country={USA}}


\begin{abstract}

In the quest to understand the climates and atmospheres of exoplanets, 3D global climate models (GCMs) have become indispensable. The ability of GCMs to predict atmospheric conditions complements exoplanet observations, creating a feedback loop that enhances our understanding of exoplanetary atmospheres and their environments. This paper discusses the capabilities of the Global Exoplanet Spectra (GlobES) module of the Planetary Spectrum Generator (PSG), which incorporates 3D atmospheric and surface information into spectral simulations, offering a free, accessible tool for the scientific community to study realistic planetary atmospheres . Through detailed case studies, including simulations of TRAPPIST-1 b , TRAPPIST-1 e, and Earth around Sun, this paper demonstrates the use of GlobES and its effectiveness in simulating transit, emission and reflected spectra, thus supporting the ongoing development and refinement of observational strategies using the James Webb Space Telescope (JWST) and future mission concept studies (e.g., Habitable Worlds Observatory [HWO]) in exoplanet research.
\end{abstract}

\end{frontmatter}

\section{Introduction} \label{sec:intro}
Atmospheric characterization is a fundamental part of exoplanet science, as it enables the identification of key atmospheric constituents, temperature structures, and potential climate regimes , furthering our understanding of these distant worlds\cite{turbet_2016,Turbet_2018,Wolf_2017,Fauchez_2019,Suissa2020}. By analyzing the light filtered through or emitted by an exoplanet's atmosphere, we can infer the presence of molecules such as water vapor, carbon dioxide, and methane, essential for determining habitability and atmospheric dynamics  \citep{Coulombe2023_JWST,Kempton_2023}.
In the ever-expanding world of exoplanet research, the application of 3-dimensional (3D) global climate models (GCMs) has emerged as a pivotal component \citep{Rice_2014, Wolf_2019, Fauchez_2021,Perera_2023} in our quest to unravel the enigmatic climates and atmospheres of these distant worlds. Exoplanets offer a tantalizing glimpse into the diversity of planetary environments that exist in our galaxy and beyond. They exhibit a vast range diversity in their characteristics, such as size, composition, and proximity to their host stars, for which the dynamic interplay between their atmospheres and the conditions on their surfaces can only be captured with 3D climate models \citep[\textit{e.g.}][and references therein]{Wolf_2017,Turbet_2018,Fauchez_2019,Komacek_Abbot2019, Komacek_2020, Suissa2020, Fauchez_2021,sergeev_thai}. These models resolve and/or parameterize principles of fluid dynamics, thermodynamics, radiative transfer (RT), and other atmospheric and geological processes to represent as accurately as possible the climate state and atmospheric parameters of a given world. The development of GCMs for Earth science applications such as seen in the Coupled Model Intercomparison Project (CMIP, \citep{Eyring_2016}) has enabled their use for other planets, in the solar system such as Mars \citep{Forget_1998}, Venus \citep{Lebonnois_2010}, Titan \citep{Lora_2019} and beyond \citep{Way2017,turbet_2016,Fauchez_2019,Suissa2020}. 

One of the most significant advantages of 3D global climate models with respect to their simpler 1-dimensional (1D) counterparts \citep{Kopparapu_2013,Lincowski_2018} is their capacity to represent atmospheric dynamics, clouds, and the day/night side contrast of tidally locked exoplanets. Particularly for clouds, where the net effect on the energy balance of the atmosphere is effectively opposite on the day versus the night side, the 3D aspect is critical. Furthermore, 3D global climate models contribute to the development of sophisticated techniques for observing and characterizing exoplanets by enabling the exploration of spatial and temporal variation. Even when the planets may not be resolved in remote observations, the averaging of the 3D aspects leads to incorrect spectra, as is shown in this work.  With advancements in telescope technology and the application of spectroscopy, it has been over two decades now \citep{Charbonneau2002} that planet spectroscopy is not only limited to our solar system. With the unprecedented capabilities of the James Webb Space Telescope (JWST, \cite{Beichman_2014}) we have moved one step forward in characterizing extra solar planet atmospheres. The sensitivity, high spectral resolution, and wavelength coverage of JWST has motivated a tremendous amount of spectroscopic studies for predicting, and now interpreting, its data. The information derived from these observations can then be compared with the predictions made by climate models, creating a feedback loop that enhances our understanding of these distant worlds. Looking to the future, with the recommendation of the NASA 2020 Decadal Survey \citep{Astro2020} to develop a $\sim 6 m$ UV/VIS/NIR space telescope, currently called the Habitable World Observatory (HWO), direct imaging of Earth-Sun twin systems is potentially in the near future \citep{theluvoirteam2019luvoir}. Two mission concepts, Habitable worlds explorer (Hab-Ex, \cite{HabEx2018}) and the Large ultraviolet, optical, IR observatory (LUVOIR, \cite{theluvoirteam2019luvoir}) are heavily leveraged for the HWO concept. As with the observations of 3D effects on planetary spectra important for JWST observations (hotter and larger planets), 3D effects are expected to be important for observations of smaller rocky exoplanets using  HWO.  Therefore, GCM simulations connected to complex spectral modeling play a pivotal role in correctly assessing the detectability and optimized characterization of exoplanets. Even prior to the building of the telescope, 3D simulations are used to study the sensitivity of the observatory to atmospheres and their spectroscopic information. Parallel development of simulations capabilities to the definition of the instrument may prevent over-or under estimation of the mission yields, and thus its success. GCMs with or without photochemistry \citep{Chen_2019,Mak_2023} and 1D photochemical models \citep{Tsai_2017} have also been employed to study at the temporal variability of exoplanet spectra, unleashing the fourth dimension. While \cite{May_2021,Fauchez_2022} have shown that temporal variation in the spectra of synthetic TRAPPIST-1e atmosphere induced by climate fluctuation would not produce a detectable signature, others such as \cite{Chen_2020} and \cite{Louca_2022} have shown for rocky and mini-Neptune exoplanets, respectively, that their chemical composition, and therefore their spectral signatures, could be significant affected by photochemistry induced by  the host star. 

The array of observational techniques that can be used to extricate the properties of these worlds strongly depends on the characteristics of the star(s) and planet(s).
Transit spectroscopy \citep{Charbonneau2002} depends on the alignment of the planet-star-observer, here, the planet crosses in front of the star relative to the observer - during this event, atmospheric materials (gasses, aerosols) absorb and scatter starlight at specific wavelengths. The strength of the transmitted signal is inversely proportional to the square of the stellar radius. Because of this, the signals of rocky exoplanets are stronger around late M-type dwarfs \citep{Fauchez_2019, Komacek_2020}. For large exoplanets transiting early M-dwarfs, e.g. WASP-96b \citep{Samra_2023}, GJ 486b \citep{Moran_2023} or WASP-39b \citep{Rustamkulov_2023}, the transit transmission spectroscopy technique is still efficient, as long as clouds or hazes do not flatten the transmission spectrum \citep{Kreidberg_2014}. Note that transmission spectroscopy is also significantly affected by stellar variability in the visible and near-infrared, which can mimic transiting exoplanet atmospheric signal \citep{Lim_2023}.

Alternatively, emission spectroscopy during the secondary eclipse can provide a powerful means by which to probe into deeper portions of the atmosphere, despite the presence of clouds and hazes \citep{Miles_2023}. Thermal phase curves can also help to constrain the presence of an atmosphere and its density by modeling the efficiency of heat redistribution between the day and night side \citep{Wolf_2019}.  Ground-based observations will also be crucial, particularly high-resolution spectroscopy—possibly coupled with direct imaging using powerful coronagraphs — that may be sensitive to other atmospheric features such as of composition, temperature inversions, and dynamic phenomena. \citep{Snellen_2015,Lovis_2017,Currie_2023}

To date, most of existing retrieval algorithms assume a spherically symmetric atmosphere and employ 1D forward models to simulate spectra and atmospheric parameters of planets with a heterogeneous 3D structure. Such assumption inevitably leads to biases as shown in \cite{Feng_2016} who has demonstrated the non-uniform flux emitted by a planet can mimic methane spectral signature. \cite{Blecic_2017} demonstrated the importance of the 3D structure of an atmosphere when computing spectra by identifying the extended region effectively probed by the retrieval of secondary eclipse data, but missed by 1D forward and retrieval models.  The feedbacks between observations and theoretical work is key when interpreting exoplanet observations. Such interpretation relies on both forward RT, and backward (retrieval) modeling. Forward modeling uses a set of physical parameters (like temperature, composition) to compute synthetic spectra. In contrast, retrievals are an inverse process, in which given observational data, retrievals use statistical methods like Bayesian inference \citep{Latouf_2023,MacDonald_2023} and Markov Chain Monte Carlo (MCMC, \cite{Speagle_2020}) sampling to deduce atmospheric properties \citep{Bonfanti_2020}. While forward models simulate physics, retrievals aim to infer the parameters from data, making them more computationally expensive and statistical. Synergies between forward modeling, retrievals, and GCMs enable the connection of the existing physical and chemical models to observations  that will be informative regarding planetary formation and evolution. 
Furthermore, \cite{Line_2016} has shown by averaging two 1D transmission spectra that the presence of clouds on only parts of the limb could mimic a high mean molecular weight atmosphere. \cite{Charnay_2015,Parmentier_2018,Lines_2018,Fauchez_2019,Komacek_2020,fauchez_thai} used 3D GCM simulations and simulated transmission spectra integrating only the atmospheric profiles at the terminator, therefore ignoring the horizontal variability across the limb. In reality, when the star light is transmitted through the planet's atmosphere, it crosses first the portion of the day side adjacent to the terminator and then crosses a portion of the night side to finally be captured by the observer. Such effect has been captured in several forward models from \cite{Burrows_2010,Fortney_2010,Dobbs-Dixon_2012} and combined forward and retrieval models \citep{Caldas_2019}, yet in many cases the impact of this limb variability in the transit spectra is relatively small except potentially for gaseous planets for which the terminator region is much larger. In the case of the mini-Neptune GJ1214 b, \cite{Caldas_2019} have demonstrated that computing spectra using only terminator atmospheric columns lead to errors greater than expected noise.

The objective of this paper is to introduce an advanced simulation tool that allows to easily ingest 3D climatological data and to compute RT spectra in a realistic manner by capturing the atmospheric diversity across the disk and at the terminator. The Global Exoplanet Spectra (GlobES) is a module of the Planetary Spectrum Generator (PSG, \cite{Villanueva2018,Villanueva2022})  with a freely accessible online and easy to operate web interface (\url{https://psg.gsfc.nasa.gov/apps/globes.php}).  GlobES and PSG can be operated in an automated fashion using scripting languages (e.g., python) via a versatile Application Program Interface (API), and the models can be also installed locally via the Docker Virtualization System. GlobES has been used in several exoplanet studies such as in \cite{Turbet_2023} to simulate secondary eclipse and phase curve observation of the exoplanets TRAPPIST-1 b \& c, in \cite{Quirino_2023} to simulate transit observation of SPECULOOS-2  and in \cite{Ostberg_2023} for direct imaging of generic exo-Earths. Recently, its use in simulating Earth as an exoplanet was demonstrated \cite{KofmanGL}.
The paper is structured as follows: In Section  \ref{sec:fundations} we explain the GlobES prerequisite of RT within PSG, in Section \ref{sec:desc} we present the various components of GlobES and how to run it. We then discuss the applications to transit observations in Section \ref{sec:transit}, followed by the applications to emission spectroscopy in Section \ref{sec:emission} and direct imaging in Section \ref{sec:imaging}. Conclusions are given in Section \ref{sec:conclusions}.

\section{Fundamentals of Radiative Transfer into PSG}\label{sec:fundations}
Before presenting the details of the GlobES module in PSG, it is important to explain several important concepts relative to how PSG computes the RT in planetary atmospheres. First, the single scattering approximation is discussed, multiple scattering is covered after, then we introduce the sampling method used for pseudo 3D calculations. Finally refraction and ray-tracing are described.

\subsection{Atmospheric Scattering Using Discrete Harmonics}\label{subsec:harmonics}
The computation of light scattering, absorbing, and refracting in an atmosphere involves solving the radiative transfer equation (RTE) to understand how light interacts with atmospheric particles, such as molecules, aerosols, and cloud particles (droplets or complex crystals). This process can be treated by considering only single scattering or, more realistically, considering multiple scattering, each requiring different computational approaches. When the optical depth of the atmosphere is small ($\tau \ll 1$), the RTE can be simulated using a single-scattering approximation \citep{Villanueva2022}. Similarly, in a highly absorbing environment but with a low small scattering albedo ($\omega \ll 1$), single scattering is also commonly applied  since a photon is more likely to get removed from the medium via absorption than being scattered. The single scattering approximation in PSG employs a combination of the single scattering approximation for diffuse solar light, the Schwarzschild equation \citep{Schwarzschild_1906} for the thermal component, and an approximated multiple-scattering term for the reflected diffuse radiances \citep{Stamnes_2000,Villanueva2022}. The radiative transfer problem is decomposed into a series of single-scattering events, iteratively accounting for multiple scattering by successive approximations. By doing so, it circumvents the need for directly solving large matrix equations or performing matrix inversions, which can be computationally intensive. Instead, the method incrementally builds the solution, enhancing computational efficiency and stability.

On the contrary, when the optical depth of the atmosphere is high ($\tau \gg 1$) and in a strongly scattering $\omega \simeq 1$ environment, light will undergo several scattering events before reaching the observer. Solving the RTE for multiple scattering is more complex and requires iterative and specialized numerical methods. A widely used numerical model for this purpose is the Discrete Ordinate Radiative Transfer (DISORT, \cite{Stamnes_2000}) model, which solves the multiple scattering problem employing the discrete ordinate method and integrates several radiative transfer and numerical methods, making the solver stable for high opacities and accurate at low polar angles. PSG integrates a modified version of DISORT, called PSGDORT \citep{Villanueva2022}, which permits to handle non-LTE regimes and a variety of observational geometries, and has been optimized to operate with a variety of spectral grids (e.g., line-by-line, correlated-k) and surface scattering models. In general, the RTE can be written as \cite{Liou_2002}:
\begin{equation}\label{eq:RTE}
    \mu \frac{dI(\tau, \mu, \phi)}{d\tau} = I(\tau, \mu, \phi) - J(\tau, \mu, \phi)
\end{equation}
where $I(\tau, \mu, \phi)$ is the radiance (or intensity) and  \textit{J} the source function, which indicates the interaction of the atmospheric material with the radiation, both, the intensity and the source terms are a function of optical depth are a function of  optical depth ($\tau$), $\mu$ is the cosine of the zenith angle ($\alpha$), and $\phi$ the azimuth angle. The treatment of the zenith dependence is handled using a pseudo-spherical geometry and the solver integrates a raytracing algorithm to compute $\mu$ across the path of light in a refractive and spherical planet/atmosphere (see Section \ref{subsec:raytracing}). In RT, the source function defines the amount of radiation emitted by a medium, taking into account both its emission and scattering properties. The source function can be expressed following equation \ref{eq:SF} \citep{Liou_2002}:

\begin{align}\label{eq:SF}
    J(\tau, \mu, \phi) =& \frac{\omega}{4\pi} \int_{0}^{2\pi} \int_{-1}^{1} I(\tau, \mu', \phi') P(\mu, \phi, \mu', \phi') d\mu' d\phi'
    + \frac{\omega}{4\pi} F^* P(\mu, \phi, -\mu_0, \phi_0) e^{-\tau/\mu_0} 
    + (1 - \omega) S(\tau) 
\end{align}
where the first term is the multiple-scattering term, the second term is the single-scattering contribution to the solar diffuse radiance, while the last term is the thermal emission source (which includes non local thermodynamical equilibrium  emissions). Within the source terms, $\omega$ is the scattering albedo, $F^*$ is the solar flux at the top of atmosphere (TOA), $P(\mu, \phi, \mu', \phi')$ is the phase function with incidence $\mu', \phi'$ and emission at $\mu, \phi$, and $\mu_0$ is the solar zenith angle (negative in the phase term since it defines incoming fluxes, while positive $\mu$ indicates outgoing fluxes). Integrating equation \ref{eq:RTE} across the path of radiation (using as a dimension the optical thickness $\tau$), the first term on the right hand side leads to equation \ref{eq:newRTE} \citep{Liou_2002}:

\begin{align}\label{eq:newRTE}
I(\tau, \mu, \phi) =& I(\tau_0, \mu, \phi) e^{-\tau_0 / \mu} 
+ \int_{\tau_0}^{\tau} J(\tau', \mu, \phi) e^{-(\tau - \tau') / \mu} \, d\tau' 
\end{align}
where the first term is the radiance at the bottom of the layer attenuated by the opacity within the layer, and the second term is the integral of the radiance across the layer. 
The integration of equation \ref{eq:newRTE} is done numerically within PSG following the path of the radiation. The calculation of the source function is done employing PSGDORT for the multiple-scattering treatment, or done employing a simplified analytical method when employing the single-scattering approximation \citep{Villanueva2022}. The single-scattering approximation can be notably fast, and can be selected in PSG by selecting $N_{\text{MAX}}$=0 in the scattering configuration, yet the user should be warned that such solutions are only valid for low scattering opacity regimes as described above.

 \subsection{Multiple scattering and the $N_{\text{MAX}}$ and $L_{\text{MAX}}$ coefficients}\label{subsec:NLMAX}

Determining the scattered radiation requires solving $I(\tau, \mu, \phi)$ by balancing out the radiation scattered, emitted and absorbed across layers and at different geometries. This is done numerically by discretizing the problem into polar angles and Legendre polynomials of the phase function. As such, in PSG $N_{\text{MAX}}$ refers to the number of discrete angles or stream pairs used to represent the angular distribution of radiances, and $L_{\text{MAX}}$ denotes the maximum order of the Legendre polynomials used to expand the phase function. Each stream pair consists of an upward and a downward direction. The total number of streams is thus $2N_{\text{MAX}}$. To control the number of streams, three factors have to be considered

\begin{itemize}
    \item \textbf{Angular Resolution:} The more stream pairs (higher $N_{\text{MAX}}$), the higher the angular resolution. This means that the RTE is solved for a larger set of angles, allowing for a more detailed and accurate representation of the angular distribution of radiance and anisotropy of the radiation.
    \item \textbf{Accuracy:} Higher values of $N_{\text{MAX}}$ improve the model's accuracy, especially in cases where anisotropic scattering is significant. 
    \item \textbf{Computational Cost:} Increasing $N_{\text{MAX}}$ also increases the computational cost greatly ($\propto N_{\text{MAX}}^3$), as for each stream the whole radiative transfer scheme has to be run separately.
\end{itemize}

The phase function is characterized as following via a series expansion of Legendre polynomials with $L_{\text{MAX}}$ terms:

\begin{equation}\label{eq:Legendre}
P(\mu, \phi, \mu', \phi') = \sum_{l=0}^{L_{\text{MAX}}} b_l P_l(\cos \Theta)   
\end{equation}

where $P(\mu, \phi, \mu', \phi')$ is the phase function, $P_l$ are the Legendre polynomials, $b_l$ are the expansion coefficients and $\Theta$ is the scattering angle between the incident and scattered directions. Similarly to $N_{\text{MAX}}$, three considerations should be made when decided on the $L_{\text{MAX}}$ values.

\begin{itemize}
    \item \textbf{Phase Function Representation:} The higher the value of $L_{\text{MAX}}$, the more terms are included in the expansion, allowing for a more accurate representation of the phase function. This is particularly important for accurately modeling highly anisotropic scattering \citep{Wiscombe_1977,Nakajima_1988}.
    \item \textbf{Scattering Accuracy:} A higher $L_{\text{MAX}}$ enables the model to capture finer details in the scattering phase function, improving the overall accuracy of the RT calculations.
    \item \textbf{Computational Cost:} Similar to $N_{\text{MAX}}$, increasing $L_{\text{MAX}}$ also raises the computational burden (but less than when increasing $N_{\text{MAX}}$) due to the additional terms in the phase function expansion. The solver incorporates at least $L_{\text{MAX}} = 2N_{\text{MAX}}$ terms in the solution. When $L_{\text{MAX}}$ exceeds this limit, PSG employs the delta-M \citep{Wiscombe_1977} and IMS \citep{Nakajima_1988} methods, that enable the handling of narrow phase functions at minimum computational cost.
\end{itemize}

The implementation of light scattering computations using PSGDORT involves selecting appropriate values for $N_{\text{MAX}}$ and $L_{\text{MAX}}$ based on the desired accuracy and computational resources. For single scattering,  $N_{\text{MAX}}$ should be set to 0, while for multiple scattering, a higher number of stream pairs ($N_{\text{MAX}}$) and higher Legendre terms ($L_{\text{MAX}}$) improves the solution's angular and phase function accuracy. By carefully choosing these parameters, the RT solver can effectively model the complex interactions of light with atmospheric particles, providing insights into various atmospheric phenomena and aiding remote sensing applications.

\subsection{Radiative Transfer Sampling}\label{subsec:sampling}

Simulations that encompass the whole or parts of the planetary disk require sub-sampling in order to properly capture the diversity in incidence and emission angles.  This is in contrast to observations where the field of view  (FOV) is much smaller than the object disk (e.g., nadir, limb, occultations, looking-up observations) where a single set of geometry parameters is typically sufficient for performing the RT calculation. Issues with RT sampling arise when the FOV samples a broad range of illuminations and surface properties (e.g., the FOV is comparable and/or bigger than the object disk); in this case, one would need to compute RT simulations over different geometries, which would be then integrated to produce a single total planetary flux. By default, PSG currently defines one set of geometry parameters and one radiative-transfer calculation per simulation, yet PSG integrates an algorithm that allows to sub-sample the observable disk into sub-sections of similar incidence and emission angles. The algorithm takes hhe disk sub-sampling parameter (N) as input. N defines the number of angle bins for the incidence and emission angles, and is entered in the target section. The incidence and emission angles range from 0 to 90$^\circ$, so a N=5 would lead to a sub-division with bins of 18$^\circ$ and leading to 22 distinct radiative-transfer regions as described in the figure below. The computational requirement scales quadratically with N, so a conscious choice has to be made between accuracy and performance, with the requirements depending on the observational phase and the medium scattering properties. Inside each individual resolution element over the planetary surface, PSG performs a full RT calculation considering a pseudo-spherical refractive geometry. The contribution function of each resolution element is established numerically by the geometry module of PSG, which divides the observable disk into 500 x 500 points, and counts numerically the number of points across the different incidence/emission bins.

\begin{figure}
\centering 
\includegraphics[width=\columnwidth]{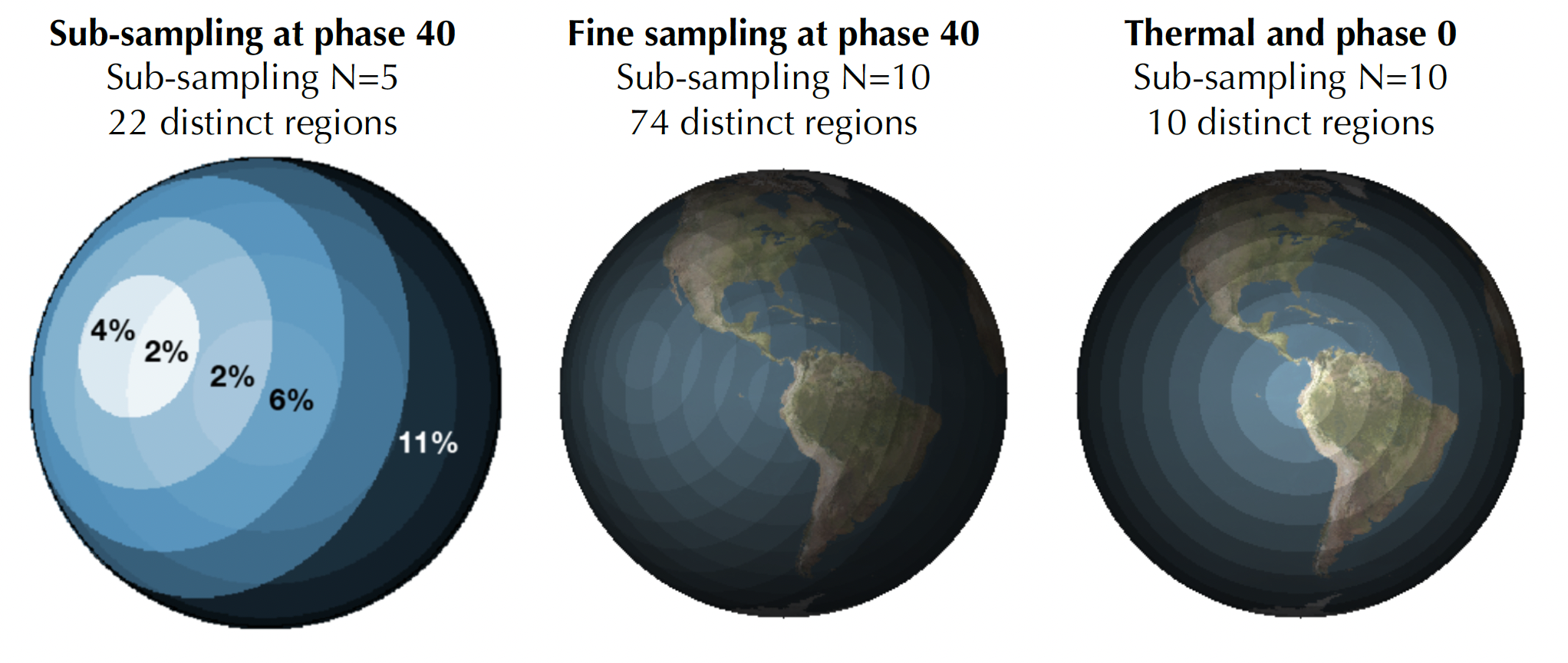}
\caption{Disk sampling methodology: when the observer beam samples a broad range of incidence and emission angles, PSG allows to divide the observable disk into sub-sections of similar incidence / emission angles. The sampling parameter (N, disk sub-sampling) defines the number of angle bins for the incidence and emission angles.}
\label{Fig:sampling}
\end{figure}

The sub-sampling approach only considers the incidence and emission angles to be changing across the disk, with the surface and atmospheric composition assumed to be homogeneous. For simulations considering heterogeneous compositions, PSG integrates the Global Emission Spectra (GlobES) module (see section \ref{sec:desc}), which allows to ingest climatological planetary 3D data. In such cases, the RT is computed at distinct regions of the planet and aggregated according to their relative projected area. Specifically, GlobES defines quadrilateral polygons (trapezoidals) of various surface areas, depending on the location over the planet, around the points where the 1D RT calculations have been computed. GlobES then performs an orthogonal projection from the planetary curved surface to the projected plane and determines the surface area of each quadrilateral using the shoelace formula:

\begin{align}\label{eq:Schoelace}
A =& \frac{1}{2}  (x_1y_2 + x_2y_3 + x_3y_4 + x_4y_1) 
- (y_1x_2 + y_2x_3 + y_3x_4 + y_4x_1) 
\end{align}

with (x$_1$,y$_1$), (x$_2$,y$_2$), (x$_3$,y$_3$) and (x$_4$,y$_4$)  the coordinates of the vertices of the quadrilateral with x$_1$, x$_2$, x$_3$, x$_4$ representing the x-coordinates of these vertices, and y$_1$, y$_2$, y$_3$, y$_4$ representing the y-coordinates. The area $A$ is calculated by multiplying and adding these coordinates in a specific sequence, and then taking half of the absolute value of this sum. The integration of the RT computations across the planetary disk is therefore performed by integrating the 1D RT calculation inside each RT  spatial bin weighted by the area $A$ of the corresponding quadrilateral.\\

Figure \ref{Fig:380nm_10um} is included to demonstrate the different strategies to calculate reflected light as a function of the planets' phase, and presents a comparison of reflectances computed with GlobES and the single atmospheric column version of PSG.  A 3D Earth-size planet with an homogeneous surface, homogeneous atmosphere and no clouds,  at 380~nm in unit of apparent albedo (I/F, upper row) is considered.

The left  panel shows i) the GlobES RT calculations performed with the maximum spatial binning (see section \ref{subsec:general} for details on the binning process) of 200 (``\textit{GlobES binned}", triangle icons), which means that GlobES internally aggregates all the atmosphere and surface columns down to one atmospheric column over an averaged surface and ii) the  RT calculations computed from the same 3D atmosphere which has been pre-averaged, offline PSG, into one single column  and for which no sub-sampling (``\textit{no sub-sampl.}", plan lines) has been applied. This test is here to demonstrate that the binning process is correctly performed within PSG. The right panel shows the  GlobES RT calculations performed at full spatial resolution without spatial binning (\textit{GlobES}, triangle icons) and the  RT calculation performed over the 3D atmosphere which has been pre-averaged, offline PSG, into one atmospheric column and surface and for which a sub-sampling of 10 has been applied ("\textit{sub-sampl.}", plain lines). ``\textit{no atm.}", ``\textit{single scat.}" and ``\textit{multi scat.}" represent the cases computed airless, with an atmosphere assuming single scattering only, and with an atmosphere assuming multiple scattering, respectively. Both ``\textit{single scat.}" ($N_{\text{MAX}}=0$) and ``\textit{multi scat.}" ($N_{\text{MAX}}=2$) are computed with four Legendre polynomials ($L_{\text{MAX}}=4$). 
First, for both panels, we can see that the \textit{GlobES binned} and ``\textit{no sub-sampl.}" cases greatly match, demonstrating that GlobES correctly bins down the spatial resolution as desired. Similarly, we can see that the fully binned RT calculation with a sub-sampling of 10 (``\textit{sub-sampl.}") of the planet disk appropriately capture the diversity of incidence and emission angles, matching the calculation at full spatial resolution (\textit{GlobES}).

Figure \ref{Fig:380nm_10um} Panel A) also shows how different the apparent albedo is  for an airless planet, for a planet with an Earth-like atmosphere with single scattering, and for an atmosphere with multiple scattering considered. As expected, the airless case has the lowest I/F as it misses scattering from gases and it follows the Lambertian reflection as in Equation \ref{eq:Lambert} \citep{Hapke_2012}.
\begin{equation}\label{eq:Lambert}
    \Phi(g)_{Lambert} = \frac{1}{\pi}(sin(g) + (\pi -g)cos(g)
\end{equation}

We can then see that the single scattering cases (``\textit{GlobES binned single scat.}" and ``\textit{no sub-sampl. single scat.}") yield slightly higher reflectivities than the multiple scattering cases (``\textit{GlobES binned multi scat.}" and \textit{``no sub-sampl. multi scat.}") at low phase angles. This is because the Rayleigh scattering phase function has a strong forward peak, which favors forward scattering compared to multiple scattering. After several scattering events, the directional information is lost, and the light is scattered uniformly in all directions, which reduces the dominance of the forward scattering peak.
The corollary is also true for intermediate phases where the ``\textit{GlobES binned single scat.}" and ``\textit{no sub-sampl. single scat.}" single scattering cases have less light that is reflected to space compared to the ``\textit{GlobES binned multi scat.}" and ``\textit{no sub-sampl. multi scat.}" multiple scattering {cases. Note that at orbital phases near 180$^\circ$ the incidence angles are very tangential and the algorithm struggles to properly capture the radiation across the disk in this marginally illuminated state. However, as we can see in Figure \ref{Fig:380nm_10um}, panel B), this issues is solved when including the sub-sampling of the planet disk. \\

In Figure \ref{Fig:380nm_10um}, panel B), we can see that the single-scattering cases always under-estimate the multiple scattering cases, to the point to be even lower than the Lambertian surface cases (``\textit{GlobES no atm.}" and ``\textit{sub-sampl. no atm.}")  at phase angles around 80$^\circ$ where the Rayleigh single scattering phase function is at its minimum. In a multiple scattering environment, despite the strong forward and backward peak of the Rayleigh single scattering phase function, ``\textit{GlobES single scat.}" and ``\textit{sub-sample single scat.}" do not overshoot the multiple scattering cases ``\textit{GlobES multi scat.}" and ``\textit{sub-sampl multi scat.}". Here, in multiple scattering regimes, the integrated effect of many such phase functions leads to a more uniform distribution of light, thereby increasing the overall reflectance. Also in multiple scattering scenarios, photons have longer effective path lengths as they undergo numerous scattering events. The longer effective path length increases the likelihood of photons being scattered in and out of the line of sight, leading to a higher overall reflectance compared to single scattering, where the path length is shorter. Furthermore, in multiple-scattering each scattering event contributes to the reflectance, accumulating to a higher value than the single scattering scenario where each photon scatters only once. Figure \ref{Fig:380nm_10um}, panel B) therefore shows that PSG sub-sampling of the planetary disk is able to accurately match GlobES RT calculations at full spatial resolution.

Note that we have also compared and validated GLobES binning and sub-sampling modeling for spectral irradiance of the same planet at 10~$\mu m$ (not shown here).

\begin{figure}
\centering 
\includegraphics[width=\columnwidth]{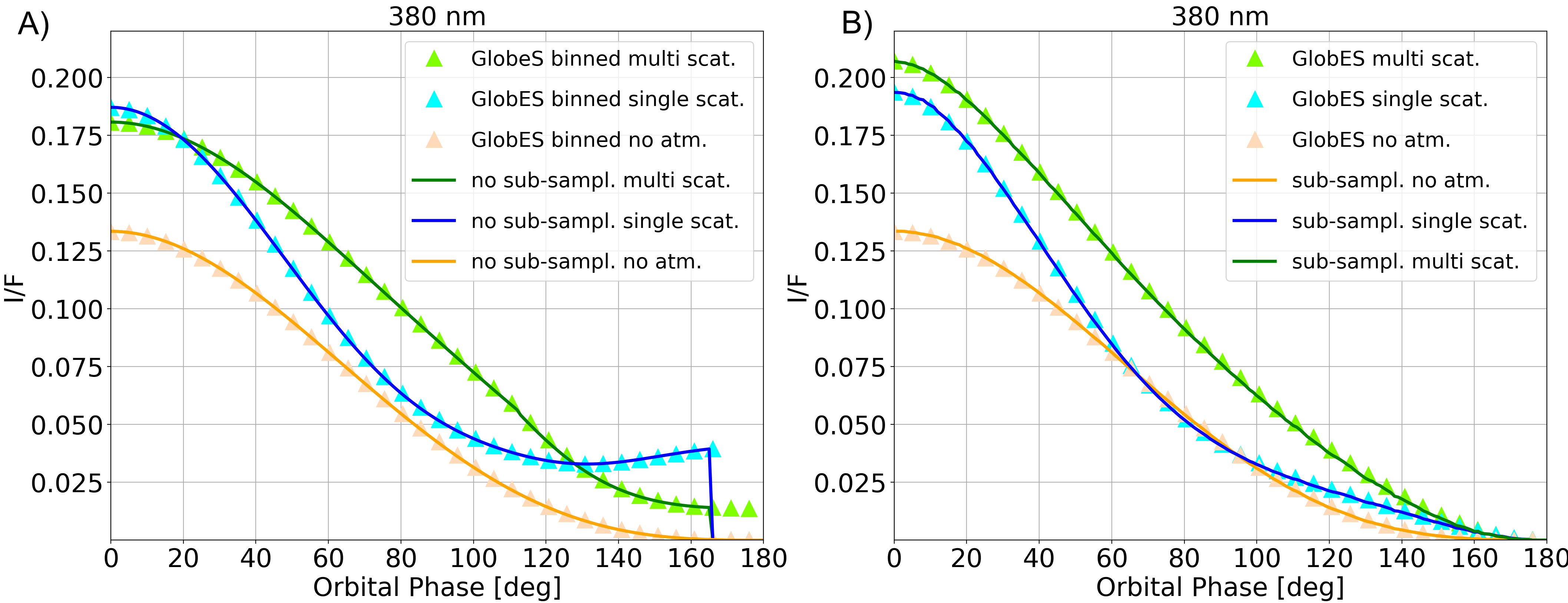}
\caption{Apparent albedo (I/F) at 380~nm  simulated for a homogeneous Earth-size planet at different orbital phases. ``\textit{GlobES binned}" represents the  RT calculations performed with the maximum spatial binning of 200, leading to a single RT calculation. ``\textit{GlobES}" represents the RT calculations at full spatial resolution (6639 RT calculations, which are then globally averaged). ``\textit{sub-sampl.}" represents the  RT calculation performed for a single 1D atmospheric column obtained from the full binning of the original 3D atmosphere with a sub-sampling of 10, while ``\textit{no sub-sampl.}" designates the absence of sub-sampling in the  RT calculation of the globally averaged 3D atmosphere. Finally, ``\textit{no atm.}", ``\textit{single scat.}" and ``\textit{multi scat.}" represent the cases computed airless, with an atmosphere assuming single scattering only and with an atmosphere assuming multiple scattering, respectively. Both ``\textit{single scat.}" ($N_{\text{MAX}}=0$) and ``\textit{multi scat.}" ($N_{\text{MAX}}=2$) are computed with four Legendre polynomials ($L_{\text{MAX}}=4$). The atmospheric composition is 1~bar Earth-like. The simulations in the left panel are fast, since only a single RT calculation is required, while, the accurate simulations on the right panel requires many simulations and are therefore more computationally expensive. Depending on the existence of strong scatterers, the surface properties and the phase, the relevance of sub-sampling and full-disk GlobES simulations will differ.
}
\label{Fig:380nm_10um}
\end{figure}

\subsection{Ray-tracing in a spherical and refractive atmosphere} \label{subsec:raytracing}

The specification of incidence and emission directions is crucial for accurately calculating radiant intensities. As summarized in the equations of sections \ref{subsec:harmonics}, \ref{subsec:NLMAX} and \ref{subsec:sampling}, all terms include a direction quantifier (\textit{e.g.}, $\mu$, f), which can be difficult to determine when the observed scene involves numerous incidence and emission angles, such as during distant planetary observations. In GlobES, the emission and incidence angles are computed by the geometry module, which are then fed to the RT module (see details in \cite{Villanueva2018,Villanueva2022}). Aside from general geometric factors, the trajectory of radiation within a spherical, refractive atmosphere is not linear but intricately curved.
Conversely, in a plane-parallel atmosphere, the incidence and emission angles across all layers are generally assumed to remain constant and are straightforwardly defined as $\mu= cos\alpha$, $\mu_0= cos\beta$. In such idealized cases, the quantity of gas interacting with all layers can be easily adjusted to the zenith value using $1/\mu$, commonly referred to as the observational ``airmass". In a spherical refractive atmosphere, however, each layer possesses unique 'entrance' and 'exit' angles, and the gas traversed within the layer follows a curved integral rather than a straight path. Chapman functions \citep[\textit{e.g.}][]{Dahlback1991} provide analytical descriptions of path lengths at each layer within a spherical atmosphere; however, they cannot account for refraction effects. Approximate curvature effects due to refraction can be derived for specific incidence angles (\textit{e.g.}, tangential) under assumptions of isothermal and simple vertical profiles \citep{Goldsmith_1963}. Nevertheless, these approximations may prove highly inaccurate when modeling realistic atmospheres, where significant variations in temperature—and consequently refractive properties—occur with altitude. The most effective method for accurately modeling and understanding refraction in a non-uniform spherical atmosphere is through numerical ray-tracing \citep{Betremieux_2013,Betremieux_2014, Villanueva2022}. In PSG, the light path is traced from the observer through the atmosphere to the surface (or surface to TOA), with adjustments made to the incidence angles at each layer interface based on the refractive properties of the layers. This approach accounts for the atmosphere's curvature, dividing each layer into 10 sub-layers for a more precise representation of refraction and sphericity. This sub-layering technique is especially important when conducting raytracing for limb, occultation, or transit geometries \citep{Villanueva_2024}.
The refraction coefficient $\eta$  in a medium is determined by the local density, composition, and the wavelength of the radiation. In PSG, the refractive coefficient for each layer (h) is calculated using the atmospheric molar mass (as a proxy for composition), the temperature and pressure of the layer (density), and the wavelength $\lambda$ as follows:

\begin{equation}
    \eta = 1 + \frac{288.15k}{1.01325}\frac{P}{T}d\eta_0(\lambda)
\end{equation}
where k is the Boltzmann constant (1.38064852e-23 [$m^2\cdot kgs^{-2}\cdot K^{-1}$]), P the pressure in bar, T the temperature in K and $d\eta_0$ is the refractive coefficient minus 1 ($d\eta_0=\eta_0-1$) at 288.15 [K] and 1 [atm]. We employ constants and equations as tabulated in \url{https://refractiveindex.info} to scale $\eta_0$ to the corresponding simulation wavelengths.
These equations are applied exclusively to wavelengths exceeding 0.15$\mu m$ - the d$\eta_0$ values are maximized at 0.15$\mu m$ for shorter wavelengths.

For H$_2$ rich atmospheres:

\begin{equation}
    d\eta_{0}^{H2} = \frac{0.148956}{180.7 - \lambda^{-2}} +  \frac{0.0049037}{92 - \lambda^{-2}}
\end{equation}

For He rich atmospheres:
\begin{equation}
    d\eta_{0}^{He} = \frac{0.1470091}{423.98 - \lambda^{-2}} 
\end{equation}

For CO$_2$ rich atmospheres:
\begin{equation}
\begin{aligned}
            d\eta_{0}^{CO2} = & \frac{6.991\times 10^{-2}}{166.175  - \lambda^{-2}} +  \frac{1.4472\times 10^{-3}}{79.609 - \lambda^{-2}} + \frac{6.42941\times 10^{-5}}{56.3064  - \lambda^{-2}} +   \frac{5.21306\times 10^{-5}}{46.0196 - \lambda^{-2}}
\end{aligned}
\end{equation}

For air (i.e. N$_2$/O$_2$) based atmospheres:

\begin{equation}
\begin{aligned}
    d\eta_{0}^{air} =&  8.06051\times 10^{-5} + \frac{2.48099\times 10^{-2}}{132.274 - \lambda^{-2}} + \frac{1.74557\times 10^{-4}}{39.32957 - \lambda^{-2}}
\end{aligned}
\end{equation}

\begin{figure}
\centering 
\includegraphics[width=\columnwidth]{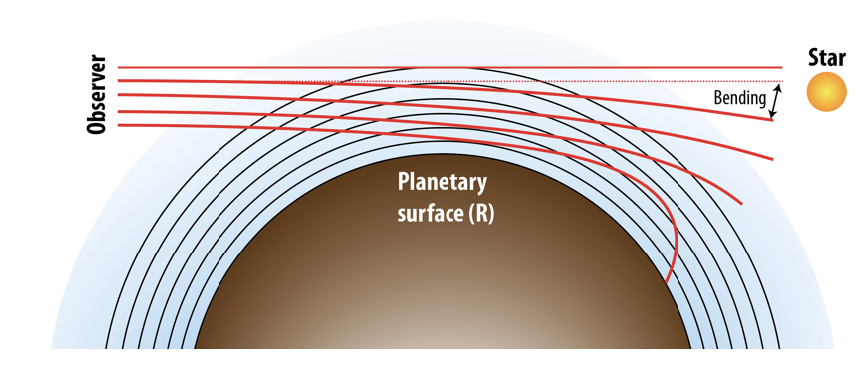}
\caption{Ray-tracing through a planetary atmosphere involves layer-by-layer variations in ray length and angles within a spherical atmosphere (gray curved lines), as refraction 'bends' the light (red traces) along its path. Depending on the degree of bending, the observer may detect radiation originating from the surface, the TOA, or a background star (\textit{e.g.}, during transit observations).}
\label{Fig:raytracing}
\end{figure}

At each sub-layer interface, the ray-path angles are adjusted using Snell's law, $sin\alpha_1 = (\eta_0 / \eta_1)sin\alpha_0$, while the gas amount integrated along the curved path within the sub-layer is calculated numerically using Simpson's 1/3 rule for integration. The ray-tracing algorithm begins at the observer's location and tracks the radiation path through the atmosphere (see Figure \ref{Fig:raytracing}). For example, in an occultation or limb geometry at an impact height $H$ for a planet with radius $R$, then the initial path angle at H$_{TOA}$ is calculated as
$sin\alpha_{TOA}=[(H+R)/(H_{TOA}+R)]sin\alpha_H$. In exoplanet transit observations, the ray-tracing algorithm in PSG must compute the ray-tracing for every atmospheric layer. Additionally, the algorithm verifies that the path does not curve beyond the size of the star or intersect the planet's surface, ensuring that stellar light reaches the observer for the corresponding layer.

\section{Description of the Global Emission Spectra (GlobES)} \label{sec:desc}

\subsection{General Considerations and Applications}\label{subsec:general}
Realistic modeling of a planetary disk or terminator region requires to consider 3D heterogeneity of a planet's atmospheres. PSG integrates the  Global Emission Spectra (GlobES, \url{https://psg.gsfc.nasa.gov/apps/globes.php}) module, that allows to ingest 3D atmospheric and 2D surface data. These includes temperature, pressure, wind, gas abundances (in unit of volume mixing ratio) and aerosols abundances (in unit of mass mixing ratio), along with surface properties produced by GCMs.

The typical output format employed by these climate models is the network Common Data Form (netCDF), which contains many climatogological and atmospheric state variables and these files can be notably large. Many of these outputs are not directly needed in a RT calculation, and a first step when operating GlobES is to extract the required variables into a smaller GlobES specific binary file. This conversion / transformation can be done using one of the GCM-specific Python scripts available on our GitHub page (\url{https://github.com/nasapsg/globes}), or employing ad-hoc scripts tailored to the user's GCM outputs. Currently available conversion scripts are for the ExoCAM \citep{Wolf2022}, Planetary Climate Model (PCM-Generic),  ROCKE-3D \citep{Way2017} and the UM \citep{Wood2014}, as well as for the Earth climate models from MERRA-2. 

Two parameter lines in the GlobES configuration file are reserved to the GCM parameters:\\
``$<$GENERATOR-GCM-BINNING$>$" and ``$<$ATMOSPHERE-GCM-PARAMETERS$>$". The former line controls the spatial binning of the GCM data to speed up the RT calculations. Higher binning means more aggregation and less PSG RT calculations. Depending on the use case on should assess the appropriate level of binning.  The GCM-PARAMETERS line specifies the format of the GCM grid and the GCM output parameters that will be ingested during the PSG RT calculation. Table \ref{tab:GCM-parameters} shows an examples of parameters that can be contained on that line, and their description.

\begin{table}
    \centering
    \caption{Detailed description of the ``$<$ATMOSPHERE-GCM-PARAMETER$>$" line in the PSG configuration file. This line provides all the information that is needed for PSG to interpret the binary GCM file. An example binary GCM file will be made available in the supplementary information.}
    \includegraphics[width=\columnwidth]{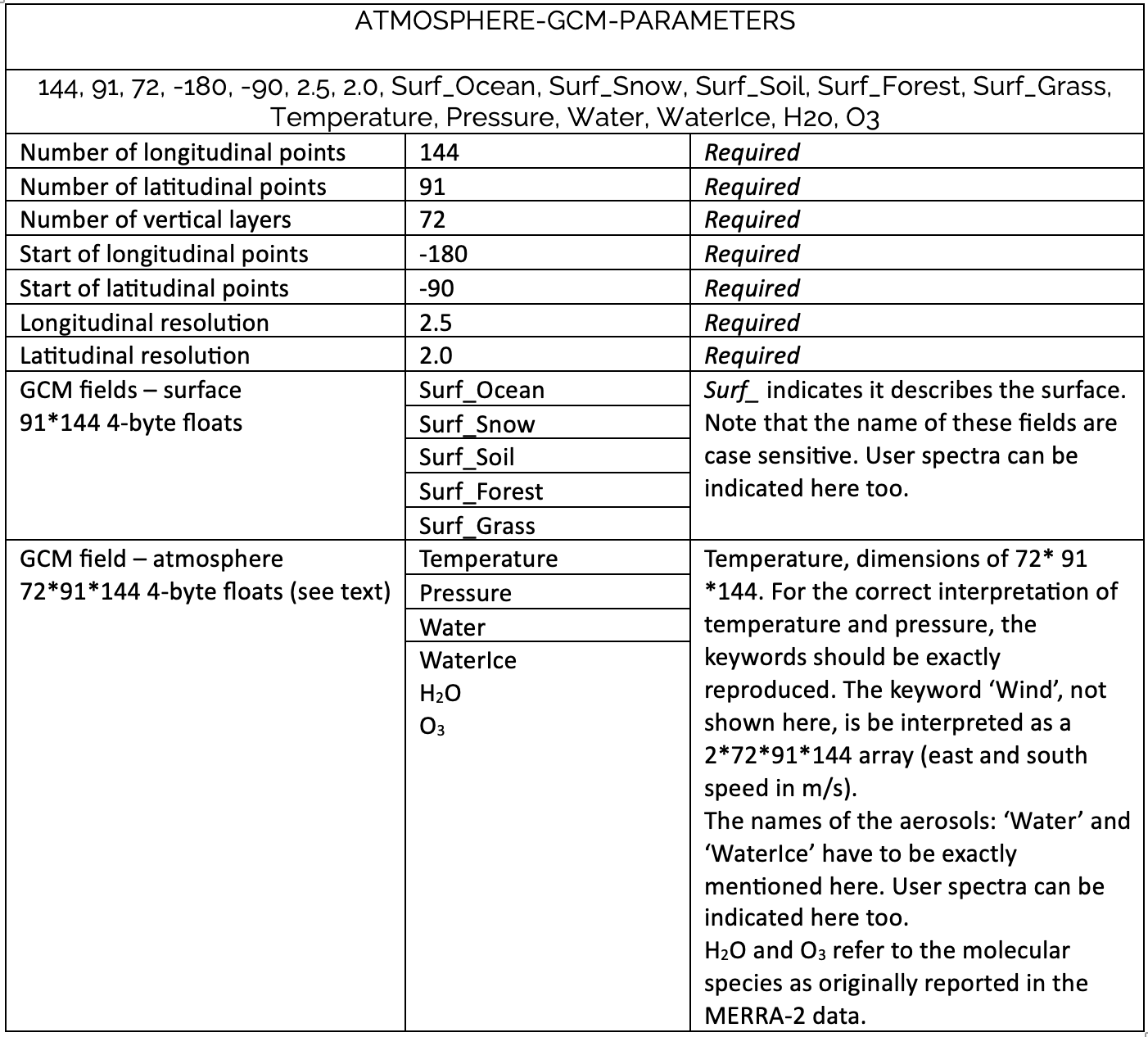}
    \label{tab:GCM-parameters}
\end{table}

\subsection{The GlobES Web Interface}\label{subsec:GUI}
The GlobES web Graphic User Interface, or GUI, is a user-friendly web application that allows to test and run spectral simulations for Earth, solar system planets and exoplanets based on GCM atmospheric outputs. A variety of templates is proposed under the ``select a source" button that allows users to visualize the 3D structure of their atmosphere and to get familiar with the options under ``Parameters of the spectroscopic 3D synthesis".
The step-by-step procedure to produce GlobES simulated spectra is detailed below: 

\begin{enumerate}
    \item Run a GCM model in your personal machine, and store the resulting 3D fields and parameters into a GCM formatted file (typically a netCDF file).
    \item Create or use one of the available Python scripts from the PSG GitHub to convert the GCM netCDF file to a PSG GCM binary file on your local machine.
    \item In the GlobES web GUI, upload the PSG GCM file by clicking on ``Load GCM data". If the upload process was successful, the site should display graphically the GCM outputs for the different variables. Check if they all look as expected at various altitude levels.
    \item Select the observing geometry (e.g., transit, disk emission/reflection) directly from GlobES webpage, or from the ``Target and Geometry" section of PSG, and verify and update the molecules and aerosols abundances and profiles to include in the simulation by visiting the "Atmosphere and Surface" section of PSG.
    \item Test your simulation with a high binning ($>$50).
    \item On the GlobES webpage, click on "Generate 3D spectra". For each position on the planet, the algorithm will update the atmospheric profiles, surface properties and geometry parameters and will run the RT. These results will be integrated employing the appropriate weights.
    \item If the spectrum looks as expected, reduce the binning appropriately for the science case.
\end{enumerate}

Figure \ref{Fig:GlobEs} illustrates how GlobES ingests GCM atmospheric data and compute spectra using as example the transit exoplanet TOI-700d ExoCAM GCM \citep{Wolf2022} simulations from \cite{Suissa2020}.

\begin{figure}
\centering 
\includegraphics[width=\columnwidth]{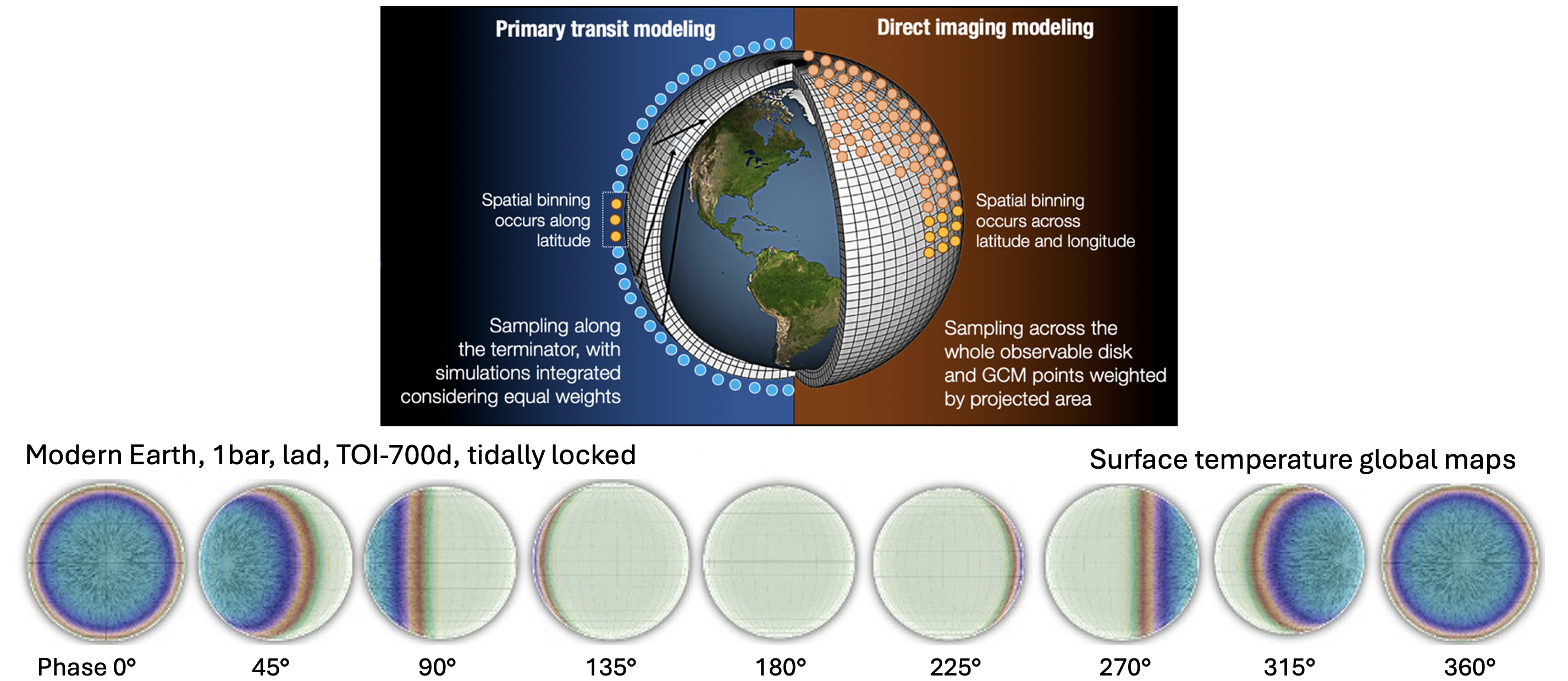}
\caption{For simulated transit observations, the GlobES app performs RT calculations across the terminator and integrates the different spectra employing equal weights. For direct imaging and secondary eclipse simulations, the algorithm performs RT simulations across the whole observable disk, and the individual spectra are integrated considering the projected area of each bin. An example of surface temperature maps at various phase angles from \cite{Suissa2020} produced from the ExoCAM GCM \citep{Wolf2022} are also shown.}
\label{Fig:GlobEs}
\end{figure}

\begin{figure}
\centering 
\includegraphics[width=10cm]{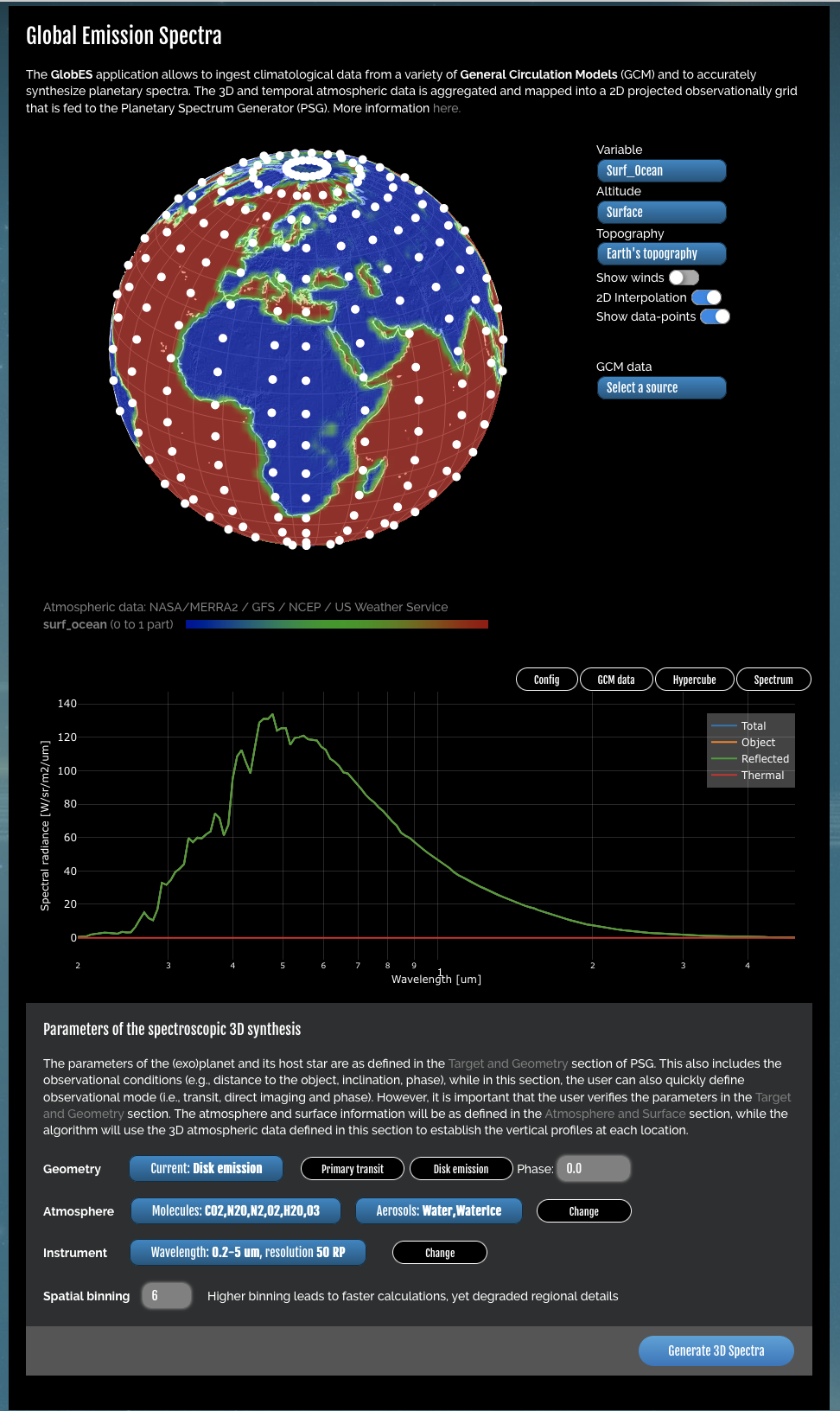}
\caption{The GlobES web interface (\protect\url{https://psg.gsfc.nasa.gov/apps/globes.php}) showing Earth spectral irradiance.  Users can load a template or upload their own GCM grid (pre-converted in a binary format, see section \ref{subsec:general}) by clicking on ``Select a source". A 0.2-5 $\mu m$ spectrum showing reflected light is computed at every white dot on the disk following the spatial binning of 6. The user can define and change the simulation parameters in the ``Parameters of the Spectroscopic 3D synthesis".}
\label{Fig:interface}
\end{figure}

\subsection{Using GlobES through the API}\label{subsec:API}

Users can also run GlobES in an automated fashion using the Application Program Interface (API, \url{https://psg.gsfc.nasa.gov/helpapi.php}). The API connects via a ``curl" commands the user-defined configuration file from a GlobES simulation to the PSG server that will compute the planetary spectra and send the output back to the user. For more details and the full list of options, see {https://psg.gsfc.nasa.gov/helpapi.php}.

The GlobES GitHub \url{https://github.com/nasapsg/globes} contains a Python script example (\url{https://github.com/nasapsg/globes/blob/main/phase.py}) that produces phase curves using an ExoCAM GCM \citep{Wolf2022} output. Lines 11 to 20 define the prescribed parameters of the simulations and lines 21 to 23 refers to the server that is going to be used for the simulations. Lines 26 and 27 convert ExoCAM netCDF outputs in GlobES binary format (.dat) while the API call is made in line 31. Users are strongly encouraged to verify the GCM conversion function used in these lines to ensure that all the relevant parameters for the simulation are included. In the API call, the script loads the GCM parameter (file@gcm\_psg.dat) to the PSG server as a new simulation (type=set). The specific parameters (wavelength, resolution, units, geometry) of the simulations are then loaded from lines 34 to 45 in to a configuration file (config.txt). A second API call is then made in line 47, which this time just update (type=upd) the existing data with the config.txt information. This process can then be made iterative across the phase space from line 50 to 57. The last lines of the scripts produce a simple plot to visualize the phase curves.

\section{Application to Transit observations}\label{sec:transit}
GlobES can be used to simulate transmission spectra based on a GCM simulation, and in the web GUI, the user should proceed as described in subsection \ref{subsec:GUI} and upload their GlobES-compatible GCM binary file. In the ``Parameters of the spectroscopic 3D synthesis" section, the Geometry should be selected as ``Primary transit" which will display a orbital phase angle of 180.0$^\circ$. If other parameters such as the atmospheric composition and instrument components need to be modified, the user will be redirected to the main PSG GUI, before coming back to the GlobES GUI. The transmission spectrum is computed by clicking on ``Generate 3D Spectra". If the spatial binning is set to 1 (no binning), then GlobES will compute one transmission spectrum per GCM grid box around the terminator (see left side of Figure \ref{Fig:GlobEs}) and will then average the spectra to obtain the planet's transmission spectrum as it would be measured by an observatory. Note that averaging the abundances and conditions of the atmosphere and running this as a single simulations would produce an unrealistic transmission spectrum that would significantly differs from the averaged spectrum computed from individual GCM gridbox spectra.
This can be seen in Figure \ref{Fig:transit} where GlobES simulated spectra are plotted for the THAI Hab1 case \citep{fauchez_thai} that uses atmospheric outputs from the ExoCAM GCM \citep{Wolf2022}. Each spectrum corresponds to a different level of latitudinal binning around the terminator, from no binning (black) to full binning (red). As the terminator is not fully cloudy, the binning aggregates cloudy and clear sky regions together, increasing the overall pixel cloudiness of the terminator region. This higher continuum level artificially decreases the amplitude of each absorption lines which would lead to a strong underestimation of their detectability. Binning the GCM outputs should therefore only be used for test purposes and the full GCM terminator resolution (binning=1) should be employed for accurate results. Using the web GUI such full calculation without any binning takes less than 1 minute to compute on the PSG servers for the spectra shown. 

\begin{figure}
\centering 
\includegraphics[width=\columnwidth]{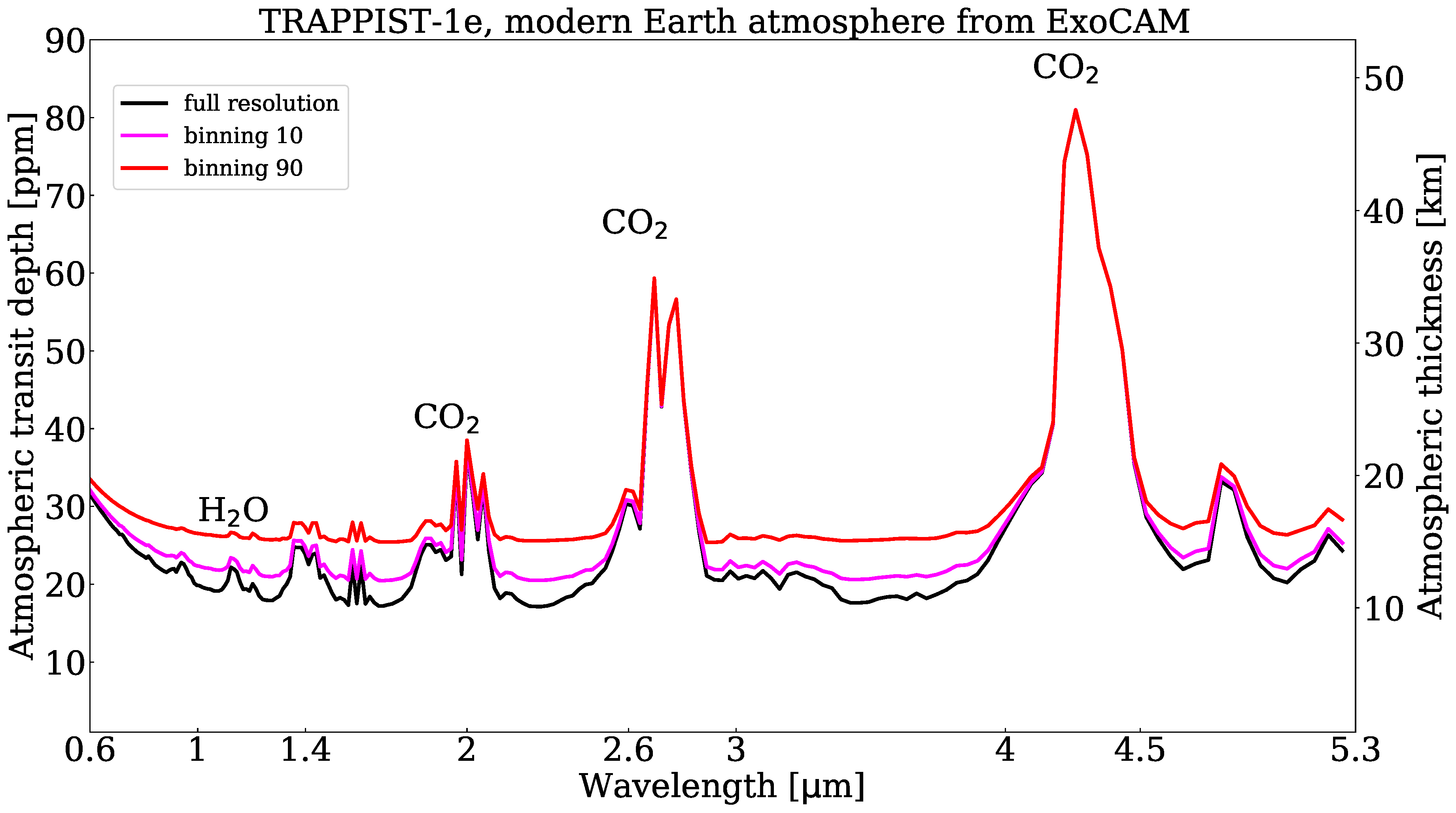}
\caption{PSG GlobES simulated transmission spectra of TRAPPIST-1 e using the THAI  \citep{fauchez_thai} Hab1 case (1 bar of N$_2$ with 400~ppmv of CO$_2$ and variable H$_2$O)  atmospheric outputs from the ExoCAM GCM using the full GCM resolution (black line), a binning of 10 terminator latitudes (pink) and a binning of 90 terminator latitudes (red, 1 average terminator profile). Binning aggregates cloudy and clear sky regions together, enhancing the overall pixel cloudiness of the terminator region, leading to a higher continuum level and shallower lines, as well as a possible loss of spectroscopic features.}
\label{Fig:transit}
\end{figure}

\section{Application to Thermal emission observations} \label{sec:emission}
Exoplanets emit in the infrared as a function of temperature, Hot Jupiters emit in the NIR \citep{Coulombe_2023} and cold exoplanets emit in the far infrared \citep{Mandell_2022}. Measuring the thermal emission of exoplanets to infer properties about their atmosphere and surface is commonly applied to hot and ultra hot Jupiters, using  Spitzer in the past \citep[\textit{e.g}][]{Knutson_2007,Stevenson2014,Stevenson2017} and now JWST \citep{Coulombe2023_JWST}. The technique can be applied to the slightly smaller hot/warm Mini Neptunes as well \citep[\textit{e.g}][]{Crossfield_2020,Kempton_2023}. The relatively small rocky TRAPPIST-1e planets have also been targeted in the infrared for characterization. TRAPPIST-1 b \citep{Greene_2023} and c \citep{Zieba_2023} have been observed with JWST’s mid-infrared instrument, MIRI, at 12.8 and 15.0$~\mu m$ with no definitive results regarding the presence of the atmosphere.
However, full F1500W phase curve observations for TRAPPIST-1 b and -1c have been assigned in JWST Cycle 2 (GO program 3077, PI Michael Gillon), for which simulated observations have been modeled using GlobES. GCMs have became indispensable tools to accurately predict, and interpret the 3D thermal structure of exoplanet atmospheres \citep{Turbet_2023} considering varying atmospheric composition and surface properties. Their capability can only be fully leveraged using a RT simulator like GlobES that is able to ingest these 3D data and produce high fidelity spectra from them \citep{Turbet_2023}.\\

In Figure \ref{Fig:emis}, we can see the GlobES simulated emission spectra for TRAPPIST-1 b, without binning and using a 1~bar CO$_2$ composition simulated with the Generic PCM GCM. Generic PCM has been used to simulate diverse exoplanetary atmospheres, from temperate rocky planets \citep{turbet_2016,Turbet_2018,Fauchez_2019} to warm mini-Neptunes \citep{Wordsworth_2011,Leconte_2013b,Charnay_2015,Charnay_2015b,Charnay_2021,Cadieux_2024}, including water-rich atmospheres for studies on runaway greenhouse and water condensation on early Earth and Venus \citep{Leconte_2013b,Turbet_2021}. It has also been applied to Martian climate studies, predicting that CO$_2$ bars are essential for sustaining liquid water on early Mars \citep{Forget_2013,Wordsworth_2013,Wordsworth_2015,Turbet_2017a}. We can see that the lowest contrast is achieved near phase 180$^\circ$ where the night side of the planet faces the observer, and that the highest contrast is achieved at phase 0/360$^\circ$ where the hotter day side of the planet is visible from the observer.
The strong CO$_2$ emission band centered at $\sim 15\ \mu m$ is clearly visible in the left panel. The difference in contrast manifests most strongly in the absorption/emission bands rather than in the continuum, therefore the contrast is lowest near secondary eclipse.
The broadband phase curve, shown in the bottom panel of Figure \ref{Fig:emis} provides additional details on where maximum and minimum thermal emission of the planets is seen. Due to the presence of an atmosphere, the hotspot is not exactly centered at the substellar point. Instead, westerly winds transport heat from the day to the night side leading to a eastward shift of the hotspot by 30$^\circ$ to phase $\sim 330^\circ$ (Turbet et al., 2024 in press). This also shifts the coldspot by the same angle. Scripts to simulate such phase curves are available at: \url{https://github.com/nasapsg/globes}.

\begin{figure}
\centering 
\includegraphics[width=\columnwidth]{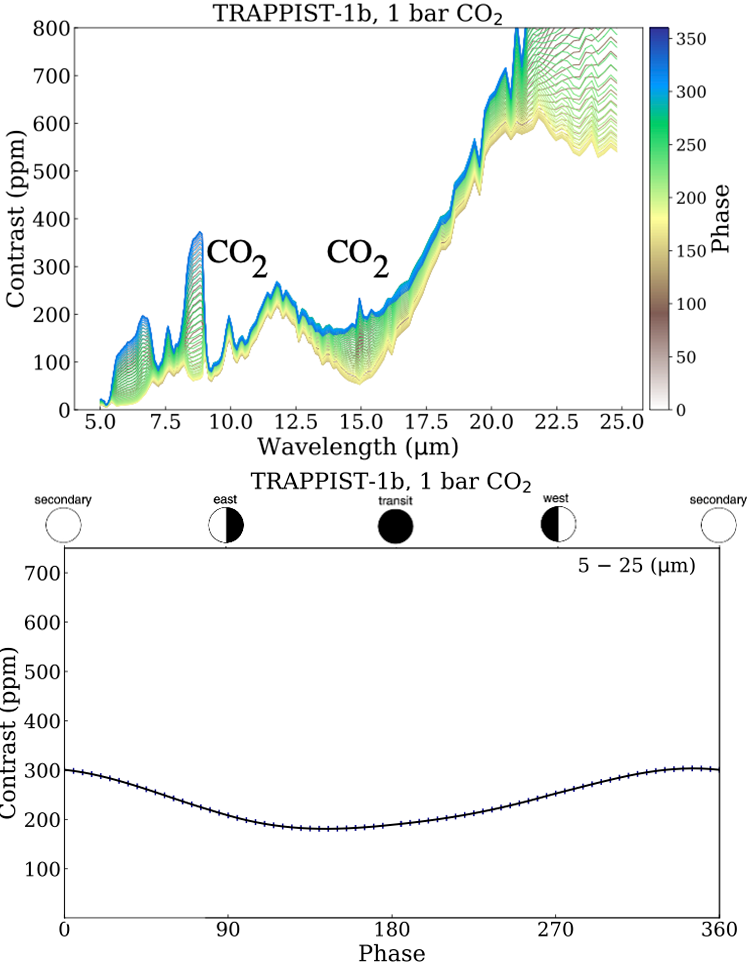}
\caption{PSG GlobES simulated emission spectra (secondary eclipse, top panel) and broadband (5--25 $\mu m$, bottom panel)  of TRAPPIST-1 b assuming a dry 1~bar CO$_2$ composition simulated with the Generic PCM GCM \citep{Turbet_2023}. Top: planet/star contrast from 5 to 25 $\mu m$ for phase angles between 0 and 360$^\circ$ with a 5$^\circ$ step increment. We can see the strong CO$_2$ absorption at $\sim 15\ \mu m$. Bottom: broadband (integrated from 5 to 25 $\mu m$) contrast as a function of the phase angle.}
\label{Fig:emis}
\end{figure}

\section{Application to Direct imaging observations} \label{sec:imaging}
\begin{figure}
\centering 
\includegraphics[width=\columnwidth]{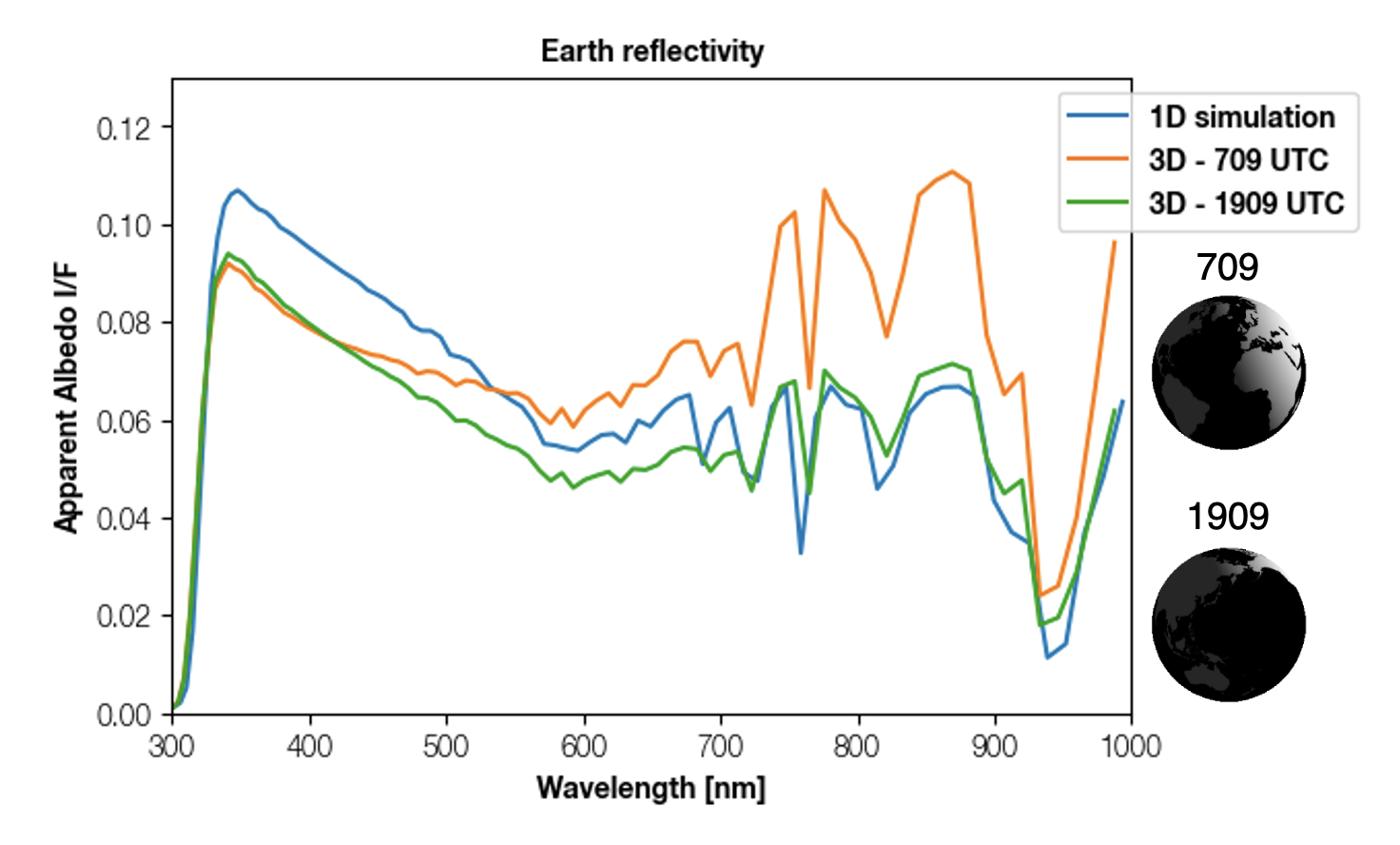}
\caption{PSG GlobES Earth reflectance spectra from 300 to 1000 nm for a 1D simulation and two 3D simulations at different times. The 1D simulation consider an average ground cover and cloud cover for the entire Earth. The 3D simulations correspond to the orientations on the right side of the figure, with universal time of the simulations indicated. Strong variations in the apparent albedo are seen depending on whether the majority of the illuminated disk is ocean or land covered surface.
}
\label{Fig:direct imaging}
\end{figure}

 GlobES enables simulations of direct imaging observations, for instance  from a HWO concept mission by ingesting 3D GCM simulations, and evaluating how the reflected light will be modulated at each of the spatial locations selected. Figure \ref{Fig:direct imaging} shows a series of simulations of Earth reflectivity, expressed in the apparent albedo of the planet (I/F), at quadrature between 300 and 1000 nm at a resolving power of 70. The atmosphere consists of the molecular species that are most important in this range (O$_2$, H$_2$O, and O$_3$), along with modeled water based liquid and ice clouds. The volume mixing ratio of H$_2$O, O$_3$, and abundance of liquid and ice clouds is varied for each location in the atmosphere. The full model details are covered in \cite{KofmanGL}. 

Three simulations are shown in Figure \ref{Fig:direct imaging}. The first considers a 1D Earth with an average ground albedo of 0.3 and average profiles of temperature, pressure and molecular abundances for the Earth. The planet is simulated to be observed at quadrature. Note that average albedo of the Earth includes the reflectivity of clouds, and is often obtained from the entire spectral range between 0.3 and 3.0 $\mu m$. Using the GlobES module, is it possible to simulate different orientations of the Earth where the albedo can be varied depending on the surface types (i.e., ocean vs vegetation) and pixel-by-pixel cloud cover can be included. The two 3D simulations show are from \cite{KofmanGL} and correspond to a quadrature view where mostly land is visible (sub-observer point Africa, 7:09 UTC ) and where mostly ocean is visible (sub-observer point in the Pacific Ocean, 19:09 UTC). For both cases the eastern half of the sphere is illuminated (also see sketches in the Figure \ref{Fig:direct imaging}). GlobES enables the study of 3D aspects in the atmosphere and 2D surface coverage. One could for instance study the variation in the vegetation red edge as a function of orientation, which is considered a biomarker \citep{Schwieterman_2018}. It is clear that a 1D simulation steps over a lot of the variation that is seen in the actual reflectivity of the Earth, as it is not possible to consider ground coverage or variations in the atmosphere (e.g. cloud coverage).  Note that patchy cloud 1D models also exist \citep{Windsor_2023}. While such models lack 3D processes that are critical for realistic cloud formation and evolution, and cannot represent the heterogeneity of the atmosphere structure and surfaces, they are an important bridge between 1D clear sky models and full 3D GCM.

\section{Conclusion}\label{sec:conclusions}
As technology and research methodologies continue to evolve, the role of 3D GCMs in Earth, solar system and exoplanet sciences is poised to be instrumental, contributing to our ever-growing knowledge of the cosmos. Indeed, specific atmospheric and surface properties of planets can only be predicted when capturing the full 3D heterogeneity of a planet, or even 4D when considering temporal variability. These 3D heterogeneities of the atmosphere and 2D variations of the surface greatly impact the resulting spectra of a planet and their integration in interpretative and retrieval models is therefore of critical importance. As presented for transit observations, an homogeneous 1D atmosphere would lead to a continuum level that is significantly higher than when considering a full resolution simulation, ultimately underestimating the real strength of the molecular features. For emission spectroscopy, not capturing 3D planetary heterogeneities would require the use of a parameterization for the heat redistribution which would significantly bias secondary eclipse observations and phase curves. Finally, for direct imaging, a 1D homogeneous atmosphere would not capture latitudinal and longitudinal dependencies, meaning that the true strength of the different atmospheric and surface variations cannot be accurately assessed \citep{KofmanGL}. Importantly, in all cases, the detectability of molecular signatures and other potential planetary markers would be masked and affected by 3D effects, and a RT code that is also able to capture such heterogeneity would be needed to accurately capture these effects. This is precisely the role of the Global Emission Spectra module of PSG. GlobES can directly ingest GCM outputs and produce planetary spectra for any kind of solar system objects and exoplanets from the smallest rocky planets to the largest, ultra hot Jupiters and for any observing geometry and instrument mode. GlobES leverages the well-tested PSG RT suite to accurately compute fluxes, transmittances, reflectances and emissivities for a wide range of planetary objects, and considering different geometries (e.g., transit, eclipse, direct imaging).

GlobES is freely available online (\url{https://psg.gsfc.nasa.gov/apps/globes.php}) on an user-friendly web interface that allows to visualize the spectra and characteristics of any planet, while also considering tailored observational and instrument settings. For automated and/or sequential runs, GlobES can also be used via an API using the PSG servers, or using the scientists own machine via the Docker virtualization system. 
Such versatile and modular spectrum generator is expected to evolve with the diversity of new exoplanet detections and with the needs of current and future space missions such as JWST, ARIEL, Roman, HWO and ground-based extremely large telescopes (ELTs).

\section{Acknowledgments}
TJF, VK, GLV and RKK acknowledge support from the GSFC Sellers Exoplanet Environments Collaboration (SEEC), which is funded by the NASA Planetary Science Divisions Internal Scientist Funding Model. The project is supported by NASA HQ-directed ExoSpec work package under the Internal Funding Scientist Model (ISFM). The authors would also like to thank the anonymous reviewer for their very helpful comments which have contributed in substantially improving the clarity of the manuscript.


\bibliographystyle{elsarticle-harv} 
\bibliography{biblio}

\end{document}